\documentclass{article}

\PassOptionsToPackage{numbers, compress}{natbib}

     \usepackage[preprint]{neurips_2020}

\usepackage[utf8]{inputenc} 
\usepackage[T1]{fontenc}    
\usepackage{hyperref}       
\usepackage{url}            
\usepackage{booktabs}       
\usepackage{amsfonts}       
\usepackage{nicefrac}       
\usepackage{microtype}      
\usepackage{amsmath}
\usepackage{bm}
\usepackage{graphicx}
\usepackage{xcolor}

\usepackage{wrapfig}
\title{Deep Reinforcement Learning for Neural Control}

\author{%
  Jimin Kim \\ 
  Department of Electrical and Computer Engineering  \\
  University of Washington\\
  Seattle, WA, 98195 \\
  \texttt{jk55@uw.edu} \\
  \And
  Eli Shlizerman \\ 
  Department of Applied Mathematics\\
  Department of Electrical and Computer Engineering \\  University of Washington\\
  Seattle, WA, 98195 \\
  \texttt{shlizee@uw.edu} \\
}

\begin{document}

\maketitle

\begin{abstract}
We present a novel methodology for control of neural circuits based on deep reinforcement learning. Our approach achieves aimed behavior by generating external continuous stimulation of existing neural circuits (neuromodulation control) or modulations of neural circuits architecture (connectome control). Both forms of control are challenging due to nonlinear and recurrent complexity of neural activity. To infer candidate control policies, our approach maps neural circuits and their connectome into a grid-world like setting and infers the actions needed to achieve aimed behavior. The actions are inferred by adaptation of deep Q-learning methods known for their robust performance in navigating grid-worlds. We apply our approach to the model of \textit{C. elegans} which simulates the full somatic nervous system with muscles and body. Our framework successfully infers neuropeptidic currents and synaptic architectures for control of chemotaxis. Our findings are consistent with in vivo measurements and provide additional insights into neural control of chemotaxis. We further demonstrate the generality and scalability of our methods by inferring chemotactic neural circuits from scratch. 

\end{abstract}

\section{Introduction}


The objective of neural control is to manipulate neural circuits to achieve a target function. The target function could take various forms. In its simplest form the function could be a change in neural activity of an individual neuron, i.e., suppression or excitation of its membrane potential. For such a target, neural control would typically be achieved by considering individual neuron and the action of the control would correspond to external stimulation. Typically the target functions are less precise and of a more implicit nature, e.g. setting the target as a profile that a population of neurons would follow. Such function would be typical in Brain Machine Interfaces (BMI) methodologies which aim to set neural population activity such that it corresponds to desired responses~\cite{sanchez2007brain,pineau2009treating,alam2016brain}. Perhaps the most implicit and challenging aim for neural control, and also the next step for BMI, would be the achievement of a target which is set to be a particular behavior. 

Setting the behavior as a target comes with implicitness which reflects the complexity of the neural control. Neural circuits consist of multiple layers that are both static (e.g. connectome) and dynamic (e.g. neural interactions, neuromodulation). These interact together in a complex and nonlinear fashion. Therefore, a reasonable approach toward first step of achieving a behavioral target could come in the form of controlling one of these layers at a time:  \textit{neuromodulation} control and \textit{connectome} control.
Neuromodulation control corresponds to modulating neural activity in existing circuits leading to target behavior by applying continuous stimuli into groups of neurons. Such control method is of a \textit{closed loop} type where it adapts to internal or external states of the circuit. On the other hand, connectome control is a control of an \textit{architecture} where particular connectivity realization facilitating the aimed behavioral target is sought.      
For both types of control, advanced control strategies effectively mapping neural states and dynamics to implicit behavioral targets are needed. 

Reinforcement learning based control appears to be the most viable option to be considered for these task since it is generically designed to optimize implicit tasks and is efficient in selection of promising control strategies when implemented as deep reinforcement learning. The approach we present is a top-down approach where we review the problems in which deep RL succeeds and then formulate the two types of neural control in terms of these problems. In particular, we formulate neural control in terms of Grid-world navigation, a typical control example in which RL showed great success. We apply our methods to a full model of the nervous system and body of \textit{C. elegans} to control both neuromodulation and the connectome to achieve aimed chemotatic behavior. We validate our results with previous findings in-vivo and show that the deep RL methodology can serve as viable neural control framework for generic neural circuits.     

\section{Related Work}
Both model-based and data-driven methods have been proposed for neuromodulation control of neural circuits. Such methods take as an input neural activity and output an external stimulation to achieve target modulated neural activity. Model-based approaches propose to make use of an underlying simplified model and establish the control to leverage both computational and analytical features of the model. For example, nonlinear control methods for coordination of spike trains in local circuits used a simplified model of neurons to infer external stimuli to adjust oscillations related to animal locomotion patterns~\cite{brown2004phase, snyder2017stability}. Model-based approaches were also shown to be effective in inference of stimuli to trigger neurons natural rhythms \cite{moehlis2006optimal} and in adding feedback to regulate oscillations in large networks \cite{orosz2010controlling, danzl2009event}.

The use of a model is advantageous since could provide guarantees for observability, controllability and stability for the proposed control. However, such approaches are limited since they require construction of a model. This is not the case in many applications in which circuits dynamics are unknown and the only available data are sampled activities of the neurons. For these scenarios, model-free data-driven neural control methods have been proposed. In particular, deep reinforcement learning methods, such as Deep Deterministic Policy Gradient (DDPG) were developed for control stimuli into multiple neurons in oscillatory spike models and showed equivalent or preferable performance to model-based methods~\cite{mitchell2018control}. Another example of applying RL in neural control is of deep brain stimulation to suppress neural activity associated with epileptic seizures~\cite{panuccio2013adaptive,guez2008adaptive,pineau2009treating}. In this application, ensemble learning and batch mode Q-learning implemented an adaptive closed loop neuromodulation control to eliminate seizures. The work showed that reinforcement learning methods can achieve successful neuromodulation and also demonstrated the sensitivity of the methods to various definitions of both state and reward functions. These results warrant a generic framework for RL for neuromodulation control, which we define in our methods. 

Connectome control can be approached from model-based and data-driven perspectives as well. Model-based approaches are more common since they incorporate simplifying assumptions regarding the connectivity. For example, it was shown that the phase reduction method can be used to optimize the coupling matrix between a pair of limit-cycle oscillators to synchronize their activity~\cite{shirasaka2017optimizing}, in addition, connectivity within antenal lobe in insects was inferred by imposing model constraints~\cite{shlizerman2014data}. Approaches for data-driven connectome control are more scarce due to limited knowledge of the connectivity of many circuits and even more so as the application of connectome modulation requires advanced biological means. Recent progress in neuroimaging and bioengineering suggests that such capabilities are becoming available. Connectomes of several organisms have been fully or partially resolved and many smaller circuits have been mapped~\cite{white1986structure,ignell2005neuronal,shih2015connectomics,franconville2018building,xu2020connectome}. For connectome modulation, until recently, the available tools were limited to ablations of neurons or synapses in the connectome~\cite{chalfie1985neural,bargmann1993odorant,hara2001genetic,chung2009femtosecond}. However, recent optogenetic methods and methods to genetically edit the connectome have been proposed and show the promise of synthesizing neural circuits with a variety of architectures~\cite{rabinowitch2015engineering, rabinowitch2020repairing, rakowski2013synaptic, shlizerman2019driving, rabinowitch2019would,sinnen2017optogenetic}.

Notably, while connectome control would correspond to typically discrete modulation, it would nevertheless require advanced nonlinear dynamic control methods rather than static methods. The reason is that neural activity supported by the connectome and its relation to the target function are nonlinear dynamic processes~\cite{kopell2014beyond}. The complexity is evident from \textit{C. elegans} nervous system which connectome was fully mapped, however, ablations of the connectome cause outcomes that are difficult to predict without examination of neural activity and behavior~\cite{liu2018functional, kim2019neural, kim2019whole}.

As the complexity of data-driven neural control becomes daunting, deep reinforcement learning based control strategy stands out as the most plausible and general candidate to address these complexities. There have been number of simple yet successful control problems in the field of deep RL that can be reformulated as either neuromodulation control or connectome control. \textit{Grid-world} navigation is one of such examples which shares several aspects with neural control in which its target function can be illustrated as an episodic event. In the methods section below, we develop this analogy further by employing grid-world like learning environment as a template for building neural control state, action and reward functions.

\section{Methods}

\textbf{\textit{Agents in Reinforcement Learning.}} The goal of the agent in reinforcement learning is to learn the best \textit{action} given the \textit{environment} to maximize the \textit{scalar rewards}. This interaction of the agent and the environment is formalized as a Markov Decision Process (MDP), or as a tuple ${[S,A,T,r,\gamma]}$ governed by 
\begin{equation*}
T(s, a, s') = P[S_{t+1} = s' | S_{t} = s, A_{t} = a]
\end{equation*}
where ${S}$ is the finite set of states, ${A}$ is the finite set of actions and $T$ is the stochastic transition function from state ${s}$ to ${s'}$ given the action ${a}$. ${r(s,a)}$ is the scalar reward function being obtained at time ${t+1}$ given the state ${s}$ and action ${a}$ with a discount factor ${\gamma \in [0,1]}$. For each state ${S_{t}}$ at time ${t}$, the selection of the action is given by a policy ${\pi(s, a)}$ which maps each state with particular action. The discounted reward is defined as ${R_{t} = \sum_{k = t}^{T}\gamma^{k- t}r_{k}(s_{k},a_{k})}$ and corresponds to the discounted sum of future rewards gained by the agent. The goal of the agent is to learn the optimal policy ${\pi^{*}(s, a)}$ which maximizes ${R_{t}}$ for all ${t}$. 

An effective way to learn such policy is to formulate it as the function of the discounted reward to each of the state-action pairs, i.e., q-values, 
\begin{equation*}
q^{\pi}(s,a) = E_{\pi}{[R_{t}|S_{t}=s,A_{t}=a]}
\end{equation*}
By setting ${\pi = \pi^*}$, one can then learn the optimal policy ${\pi^{*}}$ by learning the q-values for every state-action pairs. Once the q-values are learned, the optimal policy at a given state is to choose the action with highest value with probability ${(1-\epsilon)}$ where ${\epsilon}$ is the exploration factor. Such selection of the action is called ${\epsilon}$-greedy action.

\textbf{\textit{Deep Q-Learning (DQN).}} 
Learning q-values becomes intractable for large state and action spaces. DQN was introduced to address this issue by estimating q-values with deep neural networks \cite{mnih2015human}. In DQN, the agent is a multi-layered deep neural network which takes ${S_{t}}$ as the input and outputs q-values for possible actions. The DQN agent is trained as follows: For each step, the state ${S_{t}}$ is fed into the input layer and the agent outputs the action ${A_{t}}$ by following ${\epsilon}$-greedy principle. The \textit{experience} is then defined as the tuple ${[S_{t}, A_{t}, R_{t}, S_{t+1}]}$ where ${R_{t}}$ is the reward corresponding to action ${A_{t}}$ and ${S_{t+1}}$ is the new state. The experience is then stored in experience relay memory which holds the finite number of past experiences. The parameters of DQN agent are then trained using gradient descent to minimize the loss 
\begin{equation*}
ls = (R_{t+1} + \gamma_{t+1}max_{a'}q_{\overline{\theta}}(S_{t+1},a')-q_{\theta}(S_{t},A_{t}))^{2}
\end{equation*}
where ${t}$ is randomly selected time point from the experience relay memory. The training process maintains two networks: the \textit{evaluation network} with the parameters ${\theta}$ used to evaluate the q-values as well as choosing actions, and the \textit{target network} with parameters ${\overline{\theta}}$ which is a periodic copy of the evaluation network to help DQN agent to achieve stable learning of q-values. 

While DQN was a major leap in reinforcement learning, it also had several limitations leading to extensions proposed to address them. Among these extensions we use \textit{Double DQN, Prioritized experience relay} and \textit{Dueling DQN} as our DQN agent. Each of these extension brings improvements in resolving the overestimation bias of q-values, more effective sampling of past experiences and better evaluation of q-values through the decomposition into value and advantage. Since these extensions often result in synergistic effects when combined, we utilize different combinations of them to improve learning of the agent in different scenarios.   

\begin{figure}
    \begin{minipage}{0.78\textwidth}
      \centering
        \includegraphics[width=0.95\textwidth]{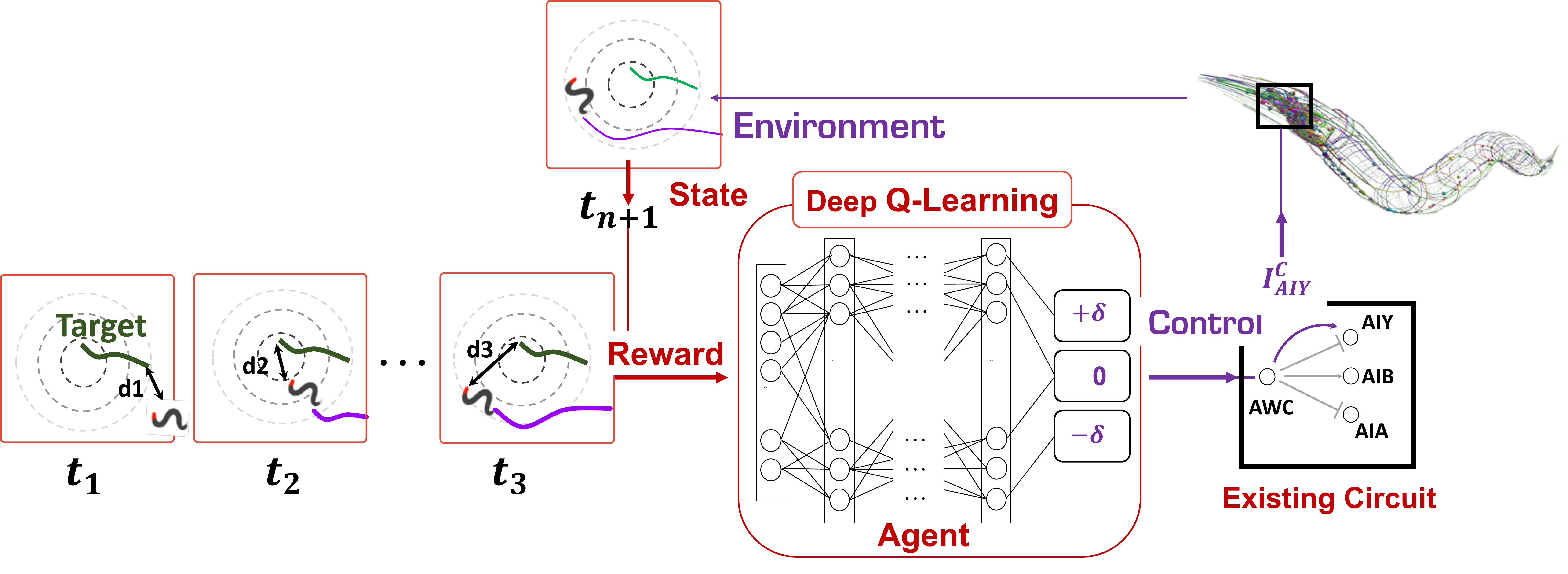}
    \end{minipage}%
    \begin{minipage}{0.22\textwidth}
      \caption{Continuous neuromodulation control for existing circuits with a Deep Q-Learning agent and grid-world like envrionment}
      \label{neuromodulation control}
    \end{minipage}
\end{figure}

\textbf{\textit{Mapping Neural Control to Grid-world Environment}}. A core idea of our method is to formulate the neural control problems as the grid-world like problem. In its simplest form, grid-world consists of spatial grids where the coordinates of each grid represent a state and an action space of size 4 allowing for going [Up, Down, Left, Right]. At ${t = 0}$, the agent is placed at the starting grid point and needs to learn the optimal policy to reach the highest reward in a shortest path while avoiding obstacles.

Adapting grid-world to the neural control problem has several advantages. Navigating gridworld is perhaps one of the most successful control tasks utilizing reinforcement learning. Deep Q-learning has shown exceptional performance in solving large and complex grid-world problem~\cite{wu2017effective}. They are also easily scalable, easy to implement, and in general, warrant convergence to good control policy. Formulating neural control tasks into grid-world like setting requires careful definition of environmental states and action spaces. We thereby define each component of control strategy for both of our neural control tasks.

\textbf{\textit{Continuous Neuromodulation  Control}}. The objective of continuous neuromodulation control is to infer control input into a set of neurons to achieve target behavior. Such behavior is often defined as sequence of spatial positions ${\bm{p}(t) = (x(t), y(t))}$ or as a function of environmental factors experienced by the neural circuits, such as sensory input ${I^{s}(t)}$ for ${t \in [t_0, T]}$. Let us define the current position and the target position at time ${t}$ as ${\bm{p}(t)}$ and ${\bm{p}^{*}(t)}$ respectively. We define the environment state and possible actions at time ${t}$ as 
\begin{equation*}
S_{t} = [d_x(t), d_y(t), I^{s}(t), I^{c}(t)], \; A_{t} = [-\delta, 0, +\delta]
\end{equation*}
where ${d_{x}(t), d_{y}(t)}$ are ${x}$ and ${y}$ components of the difference vector ${\Delta{\bm{p}(t)} = \bm{p}(t) - \bm{p}^{*}(t)}$, ${I^{s}(t)}$ and ${I^{c}(t)}$ are environmental stimulation (e.g. sensory input) experienced by neural circuit and neuromodulatory control input respectively at time ${t}$. The action space i.e. the neural control output, ${A_{t}}$ defines the incremental change to neuromodulatory control input ${\Delta I^{c}(t)}$ at time ${t}$ where ${\delta}$ is some small scalar value. Given the definitions of ${S_{t}}$ and ${A_{t}}$, we define the reward ${R_{t}} = r_{t}(s_{t},a_{t})$ as follows
\begin{equation*}
r_{t}(s_{t}, a_{t}) = \begin{cases} 
      -tanh(\alpha\Delta d(t+1)) & t < T \\
      f(I^{s}(T)) & t = T
   \end{cases}
\end{equation*}
where ${\Delta d(t+1) = |\Delta \bm{p}(t+1)| - |\Delta{\bm{p}(t)}|}$, ${\alpha}$ is a scalar that adjusts slope of ${tanh}$ function, and ${f}$ is an evaluation function that maps ${I^{s}(T)}$ to scalar. The state ${S_{t}}$ can be thought of as a grid coordinate in a grid-world at time ${t}$, where DQN agent needs to pick which direction to move to get closer to the reward (Figure~\ref{neuromodulation control}). In our formulation, these directions take the form of increasing, decreasing or not changing the neuromodulatory control input, ensuring the continuity while also keeping the action space small and discrete. The reward function is designed such that the agent minimizes the error between the target and the controlled behavior throughout the episode and maximizes the reward at the end of the episode. 

\textbf{\textit{Connectome Control}}. An alternative way to achieve aimed behavior is to modulate the architecture of the circuit. Unlike neuromodulatory control, connectome modulation is static and once set does not require the agent. Specifically, the goal of the connectome control is to infer the connectome matrix ${G^{*}}$, which achieves the target behavior ${\bm{p}^{*}(t)}$ given the sensory input ${I^{s}(t)}$ for ${t \in [t_0, T]}$.

Connectome modulation can be implemented in two forms. One can either start from an existing connectome and to make minor changes to the circuit by inserting or deleting connections, or to start from a connectome with no connections, and to attempt and infer the full circuit architecture from scratch. Both of these cases take the form of ${G^{*} = G^{0} + \Delta G}$, where ${G^{0}}$ is the initial connectome matrix and ${\Delta G}$ records changes that have been made to the initial connectome. We propose to use a DQN agent to infer ${\Delta G}$ in Grid-world like setting.

We represent ${\Delta G}$ as a function of \textit{state transition} steps similar to grid-world environment, i.e. ${\Delta G = \Delta G_{k_{f}}}$ where ${k_{f}}$ is the end step of each episode. We assign enumeration of all possible neuron pairs within the circuit to serve as transition steps such that the agent modifies a single connection at a time. The state and the action spaces are then defined as follows
\begin{equation*}
S_{k} = [\Delta G_{k}, k, m_{k}, d_{k}], \; A_{k} = [-1, 0, 1]
\end{equation*}
where ${k}$ is the step index, i.e. the neuron pair being modified, ${\Delta G_{k}}$ is ${\Delta G}$ at step ${k}$, ${m_{k}}$ are available number of insertions at step ${k}$ and ${d_{k} = k - m_{k}}$ is the distance to \textit{trapped} state at step ${k}$. Trapped state is defined as at step ${k}$, ${d_{k} < 0}$, i.e. there are more remaining insertions than transition steps, leading to ${m_{k_{f}} > 0}$. ${m_{k}}$ monitors whether the agent attempts to insert a new connection when there is no remaining insertions whereas ${d_{k}}$ monitors if there will be excess insertions by the end of the episode. The action space is a triplet selection between -1, 0 and 1 where -1 means deletion, 0 means no change and 1 means insertion of new connection with some arbitrary weight ${\delta}$. The reward function ${R_{k} = r_{k}(s_{k},a_{k})}$ is defined as follows: 
\begin{equation*}
r_{k}(s_{k},a_{k}) = \begin{cases} 
      +0.04 & m_{k+1} \geq 0 \; \text{and} \; d_{k+1} \geq 0 \; | \; a_{k}\\
      -0.75 & m_{k+1} \leq 0 \; \text{or} \; d_{k+1} \leq 0 \; | \; a_{k}\\
      f(\Delta G_{k_{f}})  & k = k_{f}
   \end{cases}
\end{equation*}
where ${f}$ is an evaluation function that maps ${G_{k_{f}}}$ to some scalar. Small positive reward is given to \textit{valid} action in which doesn't attempt to insert when ${m_{k} = 0}$ AND its insertion doesn't lead to trapped state. Negative reward is given to \textit{invalid} action which either attempts to insert when ${m_{k} = 0}$ OR leads to trapped state. The formulation is analogous to an agent navigating a grid-world with random obstacles with limited number of steps, where the grids are replaced with different realization of the connectomes~\cite{pardo2017time}. The reward function intends to guide the DQN agent to make modification to connectome at appropriate steps leading to maximum reward at the end step while avoiding either inserting too many or too few connections. Notably, connectome control introduces a vast combinatorial complexity in possible configurations of the architecture and without an effective method such as DQN, it becomes an intractable problem. e.g. 16 neurons circuit with 8 insertions corresponds to ${\sim}$ 840 billion variations. 


\textbf{\textit{Environment Modeling  Setup}}. The DQN methodology does not constrain the environment at all. As an example, we use the computational neuro-mechanical model of \textit{C. elegans} to set up our experiments. The model consists of a simulation of a full nervous system activity with the body~\cite{kim2019neural,kim2019whole}. In order to mimic the typical chemotaxis environment, we extend the model to support \textit{spatial gradient stimuli} which introduces stimulus into olfactory neurons in realistic way~\cite{itskovits2018concerted}. For each episode, we place spatial gradient stimuli on the left side of the spatial plane, and the worm on the right side. At ${t = 0}$, we initialize the worm with forward motion and simulate until ${t}$ reaches the final time point. We define the behavior for each episode to be the trajectory of the worm head ${s(t) = (x(t),y(t))}$ parametrized by time ${t}$. Here ${s(t)}$ is the function of parameters being controlled, such as external control stimulation into set of neurons or modulated connectomes. 

\begin{figure*}[t]
\centering
\includegraphics[width=1\linewidth]{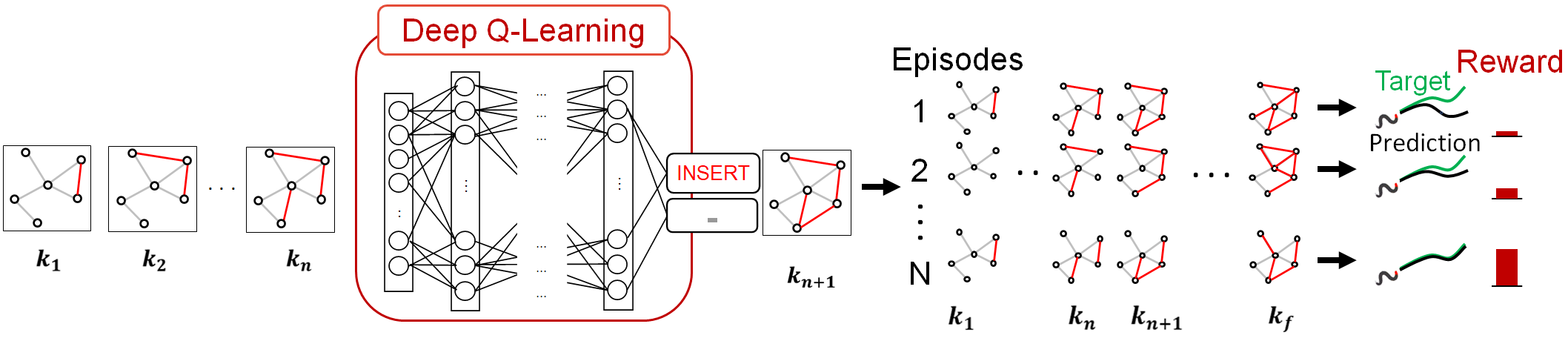}
\caption{Connectome control implemented with Deep Q-Learning agent. The agent chooses a connectome path, simulates the environment and evaluate a reward} 
\label{fig:connmod}
\end{figure*}

\section{Results}

We apply the methods described in previous section to AWC sensory circuit (i.e. AWC sensory neurons and its adjacent neighboring neurons) in \textit{C. elegans} nervous system to control chemotaxis behavior. Our results are divided into two parts for neuromodulation control and connectome control. First, we infer a continuous neuromodulatory control to emulate \textit{neuropeptidic currents} from AWC sensory neurons to AIA and AIY interneurons to obtain a baseline attraction behavior in response to AWC spatial gradient. The experiment serves as a validation of literature findings that such currents between AWC and AIA are important in achieving attraction behavior \cite{chalasani2010neuropeptide}. Next, we apply connectome control to AWC circuit to \textit{modulate the baseline chemotaxis behavior} by introducing new electrical synapses, i.e. gap junctions into the circuit. The results are validated against experimental findings which inserted synthetic gap junctions in particular locations of the AWC circuit and observed a change in chemotaxis behavior~\cite{rabinowitch2015engineering,rabinowitch2020repairing}. To further test generality and scalability of connectome control, we then expand it to \textit{infer full connectome} which achieves a desired chemotaxis behavior. For each method, we also provide comparison with other control methods in supplementary materials.     




\begin{figure}
    \begin{minipage}{0.5\textwidth}
      \centering
        \includegraphics[width=1\textwidth]{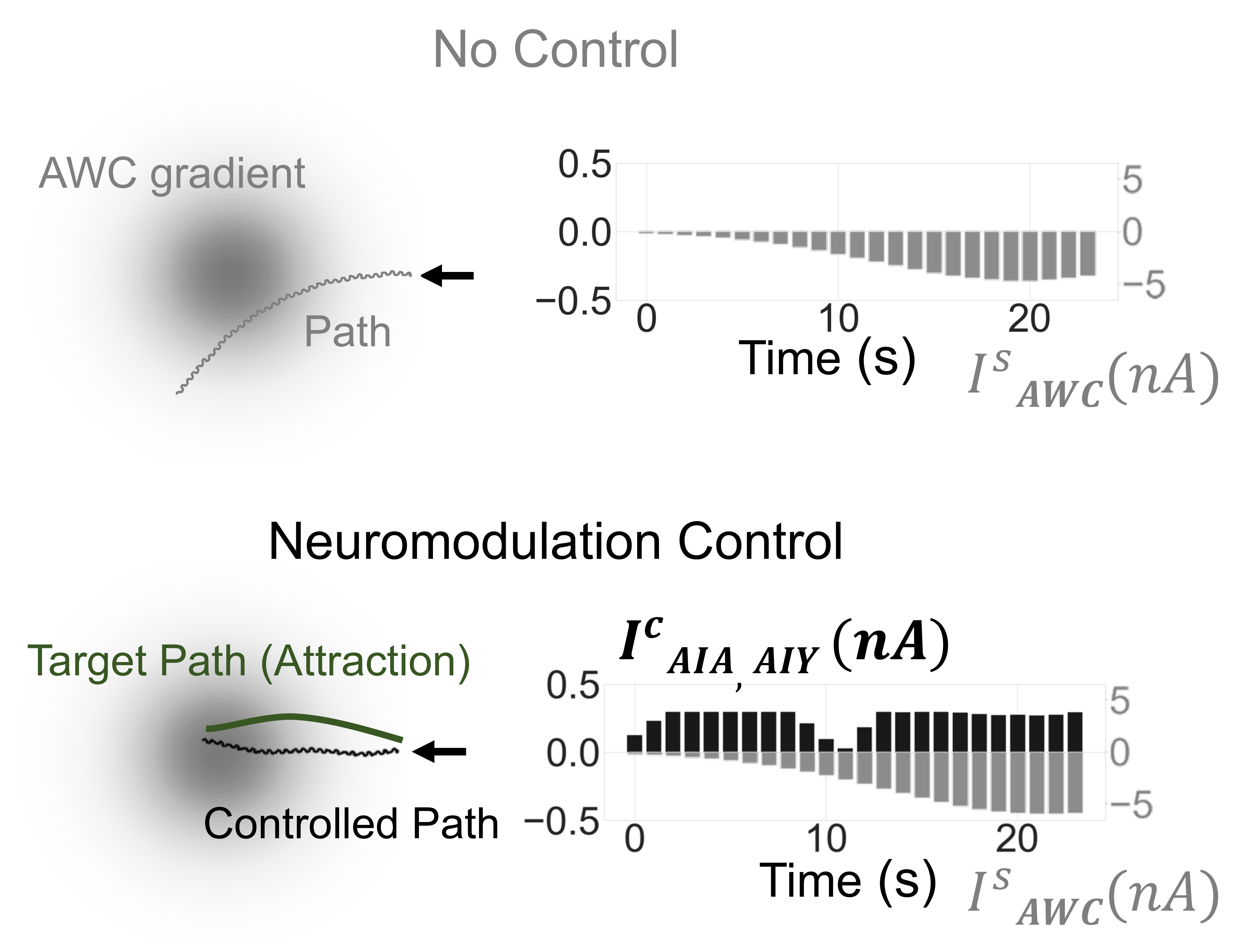}
    \end{minipage}%
    \begin{minipage}{0.5\textwidth}
      \caption{Neuromodulation control in \textit{C. elegans} chemotactic circuit to assume a target path of attraction to AWC gradient stimulus. Top: In the absence of control, the path passively passes through the gradient stimulus. Bottom: Neuromodulation control into AIA and AIY neurons changes the course such that \textit{C. elegans} is attracted to the center of AWC gradient stimulus.}
      \label{neuromodulation}
    \end{minipage}
\end{figure}

\subsection{Continuous Neuromodulatory Control}

Figure \ref{neuromodulation} top shows \textit{C. elegans} simulated behavior with no control when navigating AWC stimuli gradient. We construct AWC stimuli gradient to have a negative distribution (i.e. ${I^{s} < 0 }$) for all points throughout the space so that it mimics a typical AWC activity during chemotaxis through a odor gradient~\cite{itskovits2018concerted}. In the bottom part, neuromodulation control ${I^{c}}$ is applied to AIA \& AIY. Both control input ${I^{C}}$ and the signal from AWC gradient ${I^{s}}$ are plotted as the function of time to show the interaction between the two signals where larger amplitude of ${I^{s}}$ represents the worm getting closer to the target set at the center of spatial gradient. 

From movement paths, it is clear that neuromodulation control steers the worm toward the AWC gradient whereas the absence of control leads to worm escaping the gradient with no interaction. The control input overall maintains the range of ${\sim +0.3nA}$ but temporarily dips around ${t = 10s}$ when the worm encounters a steep gradient slope, i.e. high ${|dI^{s}/dt|}$. Remarkably, experiments have reported findings implying such adaptation of neuromodulatory currents in response to negative derivative of chemical gradient during chemotaxis \cite{itskovits2018concerted,strand1999neuropeptides,salio2006neuropeptides}. This is particularly surprising since the only constraints provided to RL agent were the lower and the upper bound for ${I^{c}}$. 
\subsection{Connectome Control}

Having an established baseline chemotaxis behavior to AWC gradient with neuromodulatory peptidic currents, we proceed to use this baseline, i.e. a closed loop system, for connectome control to introduce new connections into the circuit for novel behavioral output. We apply our method to three experiments in order to measure and test its capability to full extent. 

\textbf{\textit{Baseline behavior modulation}} In this scenario we add gap junctions in the existing AWC circuit to modulate its baseline chemotaxis behavior. Specifically RL agent is tasked with inserting gap junctions in appropriate locations in the circuit to change the behavior from \textit{baseline attraction} to \textit{repulsion}. The target is implicitly set such that for the given AWC gradient signal ${I^{s}(t)}$, the connectome that produces worm path with large curvature (i.e. steeper turn) receives higher reward. In Figure \ref{fig:insertion}, we show training results for 2 and 4 insertions scenarios (and 8 insertions scenario in supplementary material). Each column shows the learning progress in terms of number of viable candidates (i.e. the insertions that result in behavior with score higher than certain threshold) found, movement trajectory with one of such candidates, and visualization of its circuit architecture. For 2 insertion scenario, RL agent consistently finds viable configurations throughout training as evident from linear trend in number of candidates found. One of such configurations results in behavior very close to the target is connecting AWCL, AWCR and AWCR and AIAL. Strikingly, we find that such configuration is confirmed by experiment to indeed cause behavioral switch from attraction to repulsion in AWC modulated response~\cite{rabinowitch2019would}. 

\begin{figure}
    \begin{minipage}{0.6\textwidth}
      \centering
        \includegraphics[width=1\textwidth]{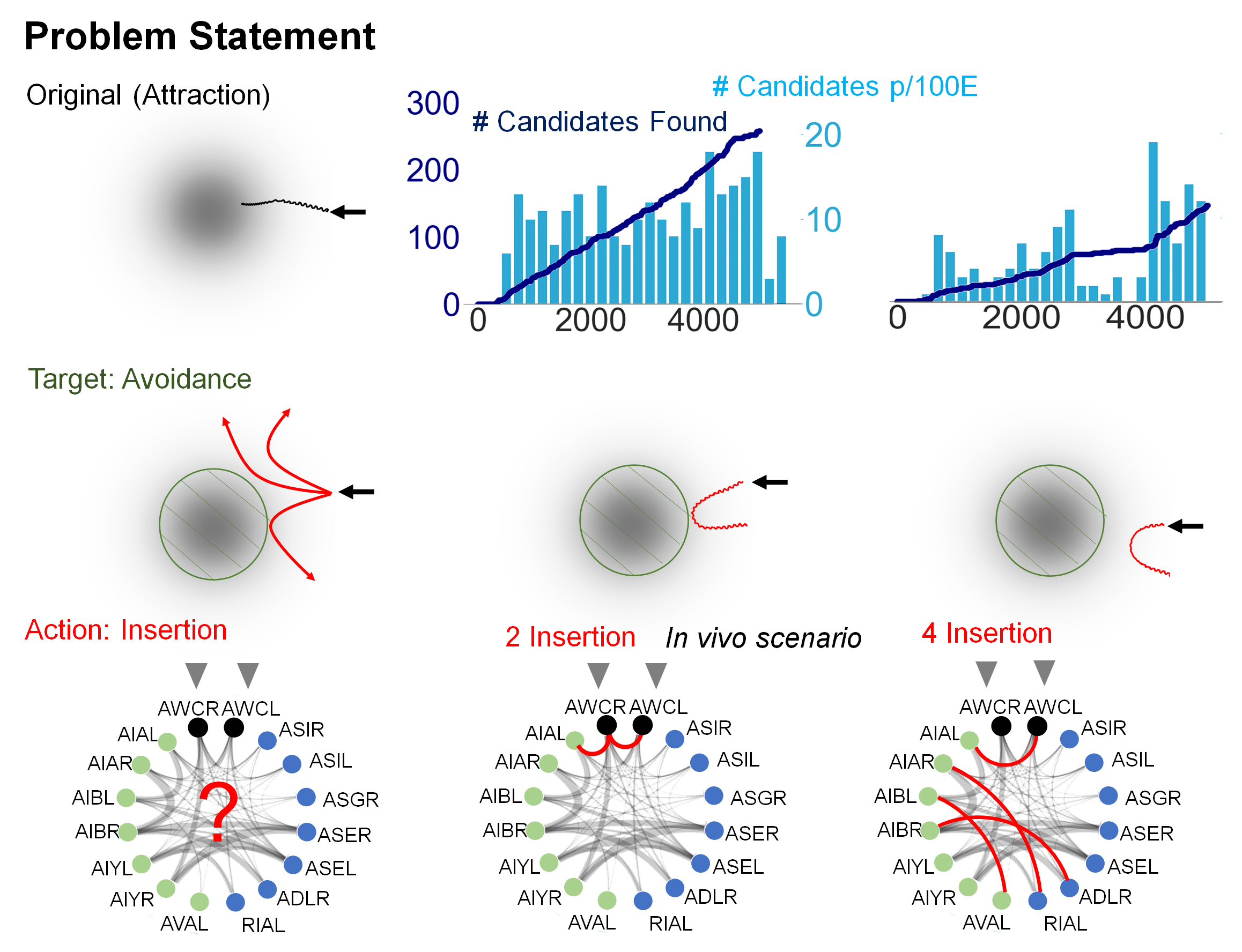}
    \end{minipage}%
    \begin{minipage}{0.4\textwidth}
      \caption{Gap junctions insertion to change baseline chemotactic behavior. Left: from top to bottom, problem statement showing original behavior before the insertion, target behavior to be achieved with insertion , and visualization of circuit subject to insertion. Middle and Right: from top to bottom, number of viable insertion configurations found as the function of training episodes, behavioral trajectory obtained from one of the candidates, visualization of the circuit corresponding to behavioral trajectory with inserted gap junctions marked in red}
      \label{fig:insertion}
    \end{minipage}
\end{figure}


Increasing the available insertions to 4 results in more challenging search as it increases both the number of possible configurations as well as the uncertainty in network activity by introducing more connections. This is indeed reflected in lower number of candidates throughout the training compared to 2 insertion scenario (Figure \ref{fig:insertion}). Interestingly enough, we notice many candidates still retain insertion between AWC and AIA, implying that the RL agent retains information on which insertions are valuable. We also notice growing number of candidates that are rather new and unexplored in literature. Such candidate is shown in the third column of Figure \ref{fig:insertion} where apart from insertion between in-vivo AIAL and AWCL, the other three connections are new and their effects are unknown. This could imply that RL agent learns how to 'engineer' the circuit such that the combined effects of multiple insertions achieve a target behavior.

\begin{figure}
    \begin{minipage}{0.6\textwidth}
      \centering
        \includegraphics[width=1\textwidth]{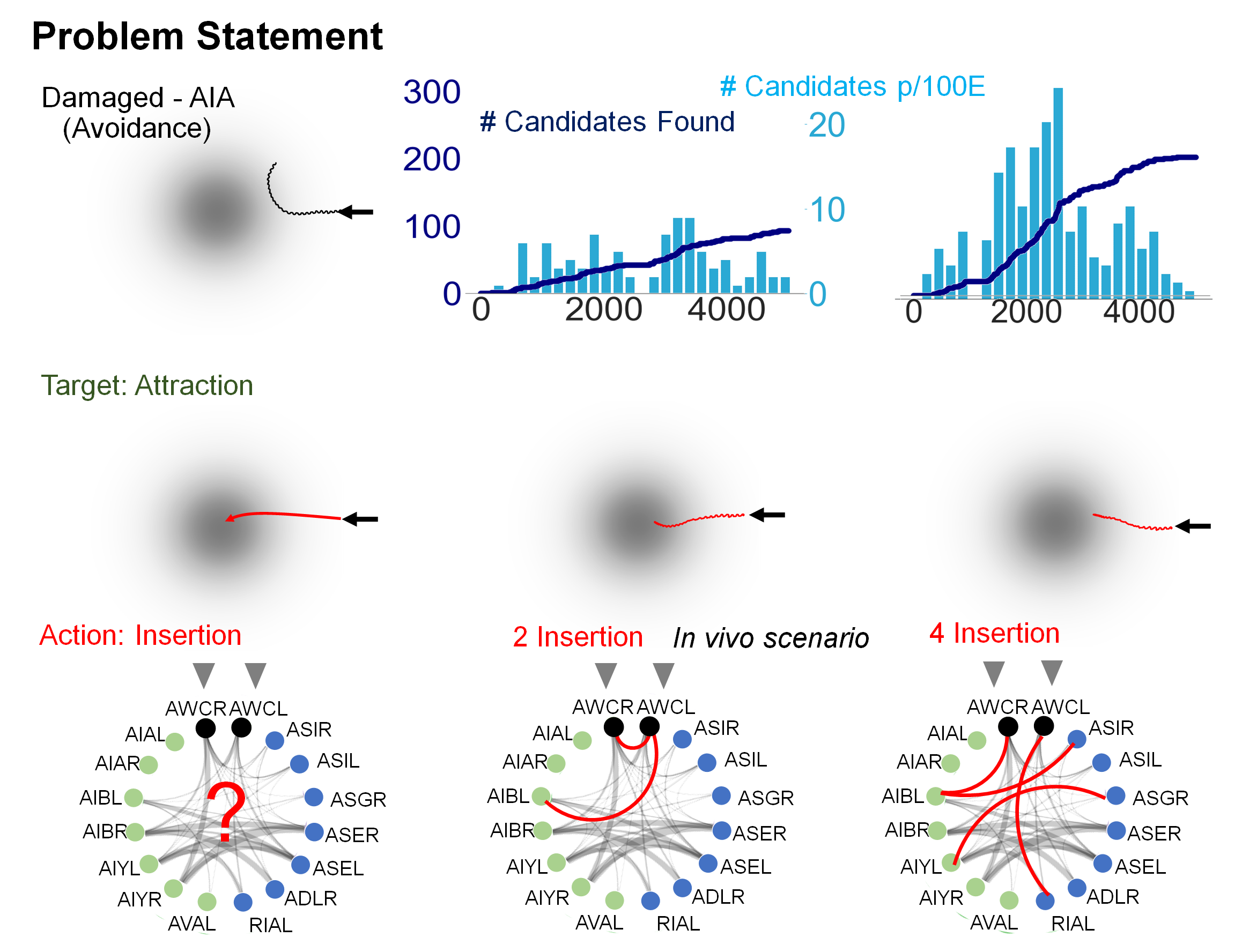}
    \end{minipage}%
    \begin{minipage}{0.4\textwidth}
      \caption{Gap junctions insertion to repair a damaged circuit (AIA ablated). Left: from top to bottom, problem statement showing chemotactic behavior with AIA ablated circuit, target behavior to be achieved with insertion, and visualization of circuit subject to insertion. Middle and Right: from top to bottom, number of viable insertion configurations found as the function of training episodes, behavioral trajectory obtained from one of the candidates, visualization of circuit corresponding to behavioral trajectory with inserted gap junctions marked red}
      \label{fig:repair}
    \end{minipage}
\end{figure}


\textbf{\textit{A damaged circuit repair}} To further test the robustness of our method, we test if we can \textit{repair} the damaged circuit and restore lost functionality by inserting new gap junctions. In particular, we ablate AIA inter-neurons from AWC circuit, leading to absence of attraction response~\cite{rabinowitch2020repairing}. The change in behavior is indeed accurately reproduced in the computational model of \textit{C. elegans} (Figure \ref{fig:repair}). We then repeat the same procedure of inserting 2, 4 and 8 gap junctions into the damaged circuit to restore the baseline attraction behavior. 

Figure~\ref{fig:repair} shows training results for 2 and 4 insertions scenarios (8 in the supplementary material) which lead to behavioral recovery. One of the candidates for 2 insertions scenario connects AWCL with AIBL and AWCR. Indeed, such configuration has been recently shown to restore the attraction behavior in AIA ablated circuit as the connection bypasses the ablated AIA neurons and lets AWC communicate directly to AIB \cite{rabinowitch2020repairing}. We notice that many of the candidates in 4 insertions scenario continue to connect AWC and AIB, further confirming the essential role of AWC-AIB coupling in behavioral recovery. Interestingly, we obtain a number of candidates that achieve a higher reward for 4 insertions than 2 insertions. One possible explanation is that since AIA is one of the most connected neurons in the circuit, it's easier for RL agent to find candidates with 4 insertions than 2 to compensate for lost connections.

\begin{figure*}[t]
\includegraphics[width=1\linewidth]{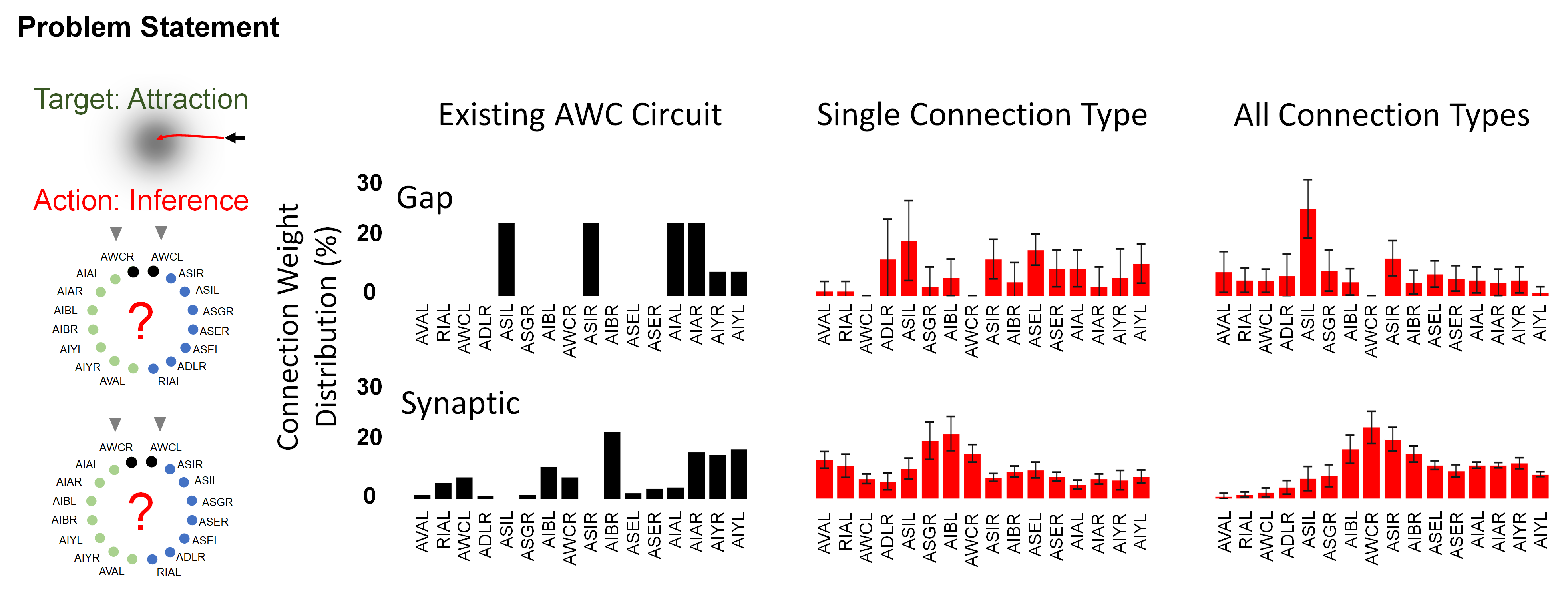}
\caption{Inferring circuit architectures to achieve target chemotactic behavior. Each column shows the connection distribution throughout the neurons in existing AWC circuit (left), inferred each connection types when other type was kept static (middle) and inferred circuits of all connection types when types are initialized with empty circuits.} 
\label{fig:inference}
\end{figure*}

\textbf{\textit{Full circuit architecture inference}} Being able to insert new connections into an existing circuit implies that the same can be done with a circuit with no connections. i.e. building the connectome from scratch. In such case, the action space of RL agent turns into discrete set of numbers where each action is \textit{synaptic weight} to be added to particular pair of neurons, allowing the agent to infer full circuit connectivity. To test such extension, we designed three experiments with same objective of inferring the circuit that can produce the baseline attraction behavior as in existing AWC circuit with neuromodulation control. Specifically, experiment 1 was tasked with inferring both gap and synaptic connectomes from scratch, whereas experiment 2 and 3 were each tasked with inferring single connection type: synaptic or gap connectome while keeping the other type of connectome intact. In order to compare the distribution of inferred connections with that of existing circuit, we set the number of insertions for each type of connectome to be the same as those in the existing connectomes. 

Figure~\ref{fig:inference} shows connection distributions between existing connectomes and inferred circuits that successfully produce baseline attraction behavior (averaged over 10 best candidates). We notice that connection distributions of inferred connectomes are generally different to those of existing connectomes even though all connection types is more similar to existing circuit than single connnection type. This implies existence of several circuit realizations to facilitate attraction. However, inferred circuits also highlight some of the hub neurons found in existing circuits, such as ASI neurons in gap connectome. It still remains an open question on why the inferred connectomes don't necessarily converge to that of existing circuit. One possible explanation is that inferred connectomes are optimized to a \textit{single} behavior type, whereas the existing circuit is for \textit{multiple} behaviors. There is a large body of evidences showing that AWC circuit partakes in sensory functions other than just olfactory~\cite{beverly2011degeneracy,kuhara2008temperature,gabel2007neural}. Thus while there may exist many circuit realizations that facilitate single behavior type as our method suggests, they might not produce sensory functions in other areas. 

In summary for both continuous neuromodulatory control and connectome control, our methods adapt grid-world like setting and were able to successfully control circuit of interest as well as to infer full circuit architecture from scratch to achieve aimed behavior.

\bibliographystyle{unsrt} 

\end{document}